# INCOME MOBILITY AND MIXING IN NORTH MACEDONIA

**Viktor Stojkoski**
*Faculty of Economics – Skopje, University Ss. Cyril and Methodius in Skopje*
vstojkoski@eccf.ukim.edu.mk

**Sonja Mitikj**
*Association for Research and Analysis - ZMAI*
sonja@zmai.mk

**Marija Trpkova-Nestorovska**
*Faculty of Economics – Skopje, University Ss. Cyril and Methodius in Skopje*
marija.trpkova.nestorovska@eccf.ukim.edu.mk

**Dragan Tevdovski**
*Faculty of Economics – Skopje, University Ss. Cyril and Methodius in Skopje*
dragan@eccf.ukim.edu.mk

***ABSTRACT***
*This study presents the inaugural analysis of income mobility in North Macedonia from 1995-2021 using the Mixing Time and Mean First Passage Time (MFPT) metrics. We document larger mobility (in terms of Mixing Time) during the '90s, with and decreasing trend (in terms of mobility) until 1999. After this year the Mixing time has been consistent with a value of around 4 years. Using the MFPT, we highlight the evolving challenges individuals face when aspiring to higher income tiers. Namely, we show that there was a noticeable upward trend in MFPT from 1995 to 2006, a subsequent decline until 2017, and then an ascent again, peaking in 2021. These findings provide a foundational perspective on the income mobility in North Macedonia.*

## 1. INTRODUCTION

The idea of upward income mobility, epitomized by the American Dream, underscores the hope that through hard work and perseverance, individuals can achieve better economic positions in societies with minimal barriers [1]. While this notion is well-discussed in the context of the United States, understanding such dynamics in other countries is equally crucial. Specifically, in North Macedonia: How long does it typically take for workers to improve their income status?

A swift progression in economic status within an individual's working life would indicate that North Macedonia's societal institutions are fostering economic prosperity and opportunity. In contrast, prolonged durations to climb the economic ladder may raise questions about the adequacy and efficiency of these societal and economic structures.

Until now, discussions about economic well-being in North Macedonia have predominantly concentrated on income inequality [2]–[4]. This dimension focuses on the disparity between the rich and the poor. However, to get a holistic view, it is vital to consider the dynamic aspects of income. These dynamics delve into how individuals or households transition between various income brackets over time, providing a more comprehensive perspective on economic fluidity.

Recognizing the importance of this perspective, our study employs state-of-the-art statistical methods to present the first estimates of income mobility in North Macedonia. Using methodologies inspired by Stojkoski [5] and Jolakoski et al. [6], we introduce metrics such as the Mixing Time, a comprehensive measure of economic mobility, and the Mean First Passage Time. The latter offers detailed insights into the duration needed for individuals to transition between different socioeconomic tiers. These measurements provide a foundation for understanding income mobility trends, enabling policymakers and researchers to ask pivotal questions. Examples of these questions include the duration required for a minimum wage worker to reach a more comfortable income level or the variations in income.

In the subsequent sections, we detail the data sources and methodologies that underpin our findings, present the results, and provide interpretations and implications based on our analysis.

## 2. METHODS AND DATA

**Measures of income mobility:** We employed two measures to understand the timeframe for Macedonian workers' income progression: the economy-wide *Mixing Time* and the individualized *Mean First Passage Time*.

1) **Mixing time:** This measure indicates how long it takes for a fresh batch of workers (like graduates) to have an income distribution similar to the current worker population. The mathematical representation involves calculating the time (in years) when the income distribution of these new entrants matches the present economy's distribution. Hence, this measure reveals the characteristic timescale of individual integration into the income distribution. We refer to Stojkoski [5] for more information about quantifying mixing in real economies.

2) **Mean First Passage Time (MFPT):** The MFPT breaks down income mobility, giving the time (in years) necessary for someone to transition between income levels (current income and target income). This measure can be further detailed to show the time to reach an income goal for specific demographic subsets, like age or gender groups. For consistency and comparability, we express starting and target incomes in terms of income distribution percentiles, ensuring our findings remain valid over time and across economies. See Jolakoski et al. [6] for more information about estimating the MFPT.

**Assumptions and Model Choice:** Our research hinges on several pivotal assumptions. Firstly, given the intricacy of income dynamics and limitations in granular data, we anchored our investigation on a null model, called geometric Brownian motion with stochastic resetting [7]. This choice is predicated on the belief that a worker's income trajectory typically involves phases of growth, punctuated by resetting events, such as job changes or losses, which revert the income to a preliminary level.

**Non-parametric Estimation Procedure:** We employed a non-parametric method, capitalizing on specific macroeconomic data points to estimate the model parameters that govern the income dynamics in North Macedonia [8], [9]. The essence of this method was to reconcile real-world income distribution data, specifically the share of income owned by the top 1%, with the theoretical stationary distribution posited by our geometric Brownian motion model. This reconciliation allowed us to extract growth and volatility parameters that resonate with the observed income dynamics of North Macedonia. Once these parameters were discerned for each year, they were then aptly converted into annual values for the Mixing Time and the MFPT. The Mixing Time was estimated using the numerical method described in [5], whereas the MFPT between any two income levels can be calculated analytically using the methods described in Jolakoski et al. [6].

We thereby emphasize that under this methodology, the yearly estimates presented are comparable across years. They measure the mobility at a given time point, under the assumption that the economy will stay in the same state as it is in the studied year. As such, we can compare the estimates in two different years and discover whether mobility has been increasing or decreasing. Finally, we point out that higher values imply lower mobility.

**Data Used in the Estimation:** Our primary datasets are twofold. First, we use the share of income controlled by North Macedonia's top 1% from 1995 to 2021. This data illuminates the concentration of income, serving as a vital proxy for discerning broader income dynamics and disparities. We took this data from the World Inequality Database [10]. Second, we collect data on resetting events, which, in our study, is proxied by figures detailing the number of individuals who either left or transitioned to new jobs within the same timeframe. This dataset underscores the frequency and magnitude of income shifts, resonating with the stochastic resetting component of our model. This data was sourced from the State Statistical Office of North Macedonia.

## 3. RESULTS

**Mixing Time Dynamics and Implications:** The Mixing Time metric, when applied to North Macedonia's income mobility from 1995-2021, reveals distinct patterns. As illustrated in Figure 1a, the '90s showcased pronounced mobility until 1997. Afterwards, mobility declined and had its lowest value in 1999 with a Mixing Time of 4.8 years. In the subsequent years, this metric stabilized around a 4-year average. While the '90s seemed to offer a dynamic economic environment, the years that followed indicate a more consistent, albeit potentially less mobile, economic landscape. Such patterns suggest that while the earlier period allowed for greater income shifts, the latter years saw more settled income trajectories, with implications for the broader socio-economic fabric of North Macedonia.

*Figure 1: Mobility in the Macedonian economy over the years.*

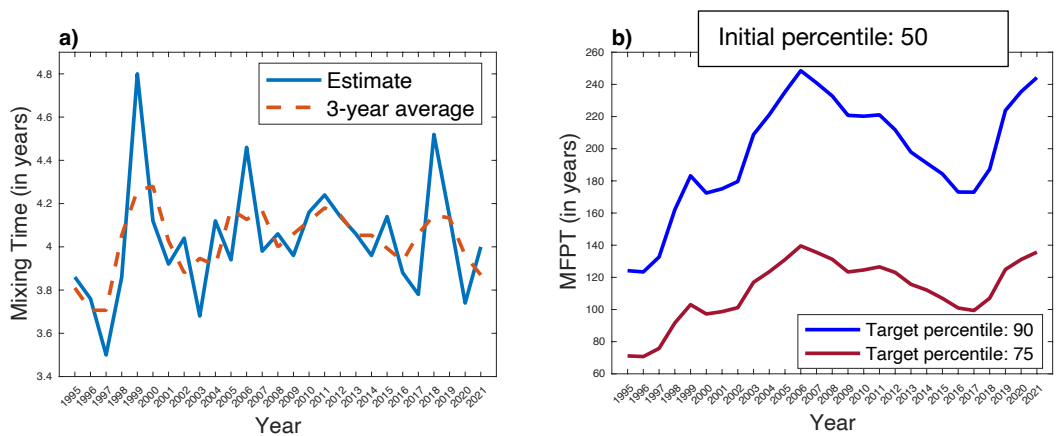

**MFPT Dynamics and Implications:** The Utilizing the Mean First Passage Time (MFPT), we can delve into the nuanced dynamics of individual income mobility in North Macedonia. Figure 1b, using the median income (50th percentile) as an illustrative starting point, highlights transitions towards the 75th and 90th percentiles. There's a noticeable upward trend in MFPT from 1995 to 2006, a subsequent decline until 2017, and then an ascent again, peaking in 2021.

The observed MFPT trends suggest that income mobility in North Macedonia experienced periods of increased fluidity and others of stagnation. The widening gap in MFPT for higher percentiles post-2017 implies greater challenges in achieving top-tier incomes. Such dynamics can impact social cohesion, economic policy direction, and workforce morale, emphasizing the need for targeted interventions to enhance equitable economic progression. We leave the exact counterfactual analysis for the reasons driving the changes in economic mobility in North Macedonia as a fruitful future research direction [11], [12].

## 4. CONCLUSION AND DISCUSSION

This study offers an illuminating view into North Macedonia's income mobility, unearthing key patterns in economic transitions. Firstly, the Mixing Time and MFPT elucidate periods of increased mobility interspersed with stasis, reflecting shifts in economic opportunities. These shifts are most likely linked to domestic economic policies or global economic trends. For instance, the drastic changes in the trend of the MFPT in 2006 and in 2017 correspond to changes in political regimes. But these years also correspond to years in which there were changes in the measurement of produced by the State Statistical Office. Thus, it is essential to delve deeper into the potential drivers of the changes in the mobility in the country. Such, counterfactual analysis is the subject of our current research.

Secondly, the widening MFPT for upper percentiles, notably after 2017, signals growing challenges in accessing the higher rungs of income, indicating possible economic stratification. Hence, it is essential to also create future projections about the economic mobility within the country.

However, our findings also come with inherent limitations. The reliance on geometric Brownian motion with stochastic resetting, although robust, may not encapsulate all complexities of real-world income dynamics. Furthermore, using data such as the share of income held by the top 1% and job transitions as proxies might introduce biases, overlooking nuanced individual experiences and regional disparities. Despite these limitations, policymakers should regard these results as foundational, urging a deeper dive into factors influencing the changes in income mobility over time. Tailored interventions promoting equitable economic opportunity are crucial, as they not only spur individual prosperity but also strengthen societal cohesion.

To sum up, our analysis, while revealing, is a starting point for measuring income mobility. It underscores the necessity for ongoing research and nimble policymaking, aiming for a comprehensive and inclusive economic landscape in North Macedonia.